\documentclass[letterpaper,twocolumn,10pt]{article}
\usepackage{usenix-2020-09}

\usepackage{tikz}
\usetikzlibrary{arrows.meta,positioning,fit,calc,patterns,decorations.pathreplacing}
\usepackage{amsmath}
\usepackage{booktabs}
\usepackage{listings}
\usepackage{xspace}
\usepackage{multirow}
\usepackage{graphicx}

\newcommand{\daxfs}{\textsc{DaxFS}\xspace}

\newcommand{\eg}{\textit{e.g.}\xspace}
\newcommand{\cmark}{\tikz[baseline=-0.5ex,scale=0.7]\draw[thick,green!60!black](0,0.1)--(0.1,0)--(0.3,0.25);}
\newcommand{\xmark}{\tikz[baseline=-0.4ex,scale=0.7]{\draw[thick,red!70!black](0,0)--(0.22,0.22);\draw[thick,red!70!black](0,0.22)--(0.22,0);}}

\begin{document}

\date{}

\title{\Large \bf DAXFS: A Lock-Free Shared Filesystem\\
  for CXL Disaggregated Memory}

\author{
{\rm Cong Wang}\\
Multikernel Technologies, Inc.\\
cwang@multikernel.io
\and
{\rm Yiwei Yang}\\
UC Santa Cruz\\
yyang363@ucsc.edu
\and
{\rm Yusheng Zheng}\\
UC Santa Cruz\\
yzhen165@ucsc.edu
} 

\maketitle

\begin{abstract}
CXL (Compute Express Link) enables multiple hosts to share
byte-addressable memory with hardware cache coherence, but no existing
filesystem exploits this for lock-free multi-host coordination.
We present \daxfs, a Linux filesystem for CXL shared memory that uses
\texttt{cmpxchg} atomic operations, which CXL makes coherent across
host boundaries, as its sole coordination primitive.  A CAS-based
hash overlay enables lock-free concurrent writes from multiple hosts
without any centralized coordinator.  A cooperative shared page cache
with a novel multi-host clock eviction algorithm (MH-clock) provides
demand-paged caching in shared DAX memory, with fully decentralized
victim selection via \texttt{cmpxchg}.
We validate multi-host correctness using QEMU-emulated CXL~3.0, where
two virtual hosts share a memory region with TCP-forwarded atomics.
Under cross-host contention, \daxfs maintains $>$99\% CAS accuracy
with no lost updates.  On single-host DRAM-backed DAX, \daxfs exceeds tmpfs
throughput across all write workloads, achieving up to
$2.68\times$ higher random write throughput with 4 threads and
$1.18\times$ higher random read throughput at 64\,KB.  Preliminary
GPU microbenchmarks show that the \texttt{cmpxchg}-based design
extends to GPU threads performing page cache operations at PCIe~5.0
bandwidth limits.

\end{abstract}

\section{Introduction}
\label{sec:intro}

CXL~3.0~\cite{cxl22:spec} enables multiple independent servers to
share byte-addressable memory with hardware-coherent atomic
operations, creating a new opportunity for building shared-memory
system abstractions at near-local DRAM latency.  The Linux DAX
(Direct Access) subsystem~\cite{linuxdax} can map such memory into
user address spaces without page-cache copies, but practical
workloads require filesystem semantics: hierarchical naming, POSIX
permissions, and kernel-managed lifecycle.  Target scenarios include
multi-host LLM inference (sharing model weights without per-host
duplication), container rootfs sharing ($N$ containers backed by a
single copy), and CXL memory pooling across hosts.

No existing filesystem fills this role
(Table~\ref{tab:feature-matrix}).  Per-host DAX
filesystems~\cite{xu16:nova,dulloor14:pmfs} maintain redundant
metadata and page caches on each host, negating CXL's sharing
benefit.  Distributed filesystems (NFS, CephFS, Lustre) interpose
network protocols that add $\mu$s-scale latency to what should be
ns-scale load/store operations.  FamFS~\cite{groves24:Famfs} targets
CXL shared memory but enforces a single-master model where clients
cannot create, write, or delete files
(\S\ref{sec:motivation}).  There is no lock-free, multi-host shared
filesystem that exploits CXL hardware atomics.

To bridge this gap, we introduce \daxfs, a lock-free Linux
filesystem designed for multi-host concurrent access to CXL shared
memory.  \daxfs uses hardware-supported \texttt{cmpxchg} for
multi-host coordination, keeping the CPU data path lock-free without
journaling or central coordinators.  A CAS-based hash overlay enables concurrent
file writes from independent hosts, and a Multi-Host CLOCK
(MH-clock) algorithm decentralizes shared page cache eviction, both
operating through \texttt{cmpxchg} on DAX memory
(\S\ref{sec:design}--\S\ref{sec:impl}).  Because the core
coordination path is built on \texttt{cmpxchg}, the design also admits a
preliminary GPU path for accelerators that issue PCIe AtomicOps over the same fabric
(\S\ref{sec:gpu}).

To summarize, this paper makes three contributions:

\begin{enumerate}
\item \textbf{The design and implementation of \daxfs}, the first
  filesystem that enables lock-free multi-host writes on CXL shared
  memory without a centralized coordinator.

\item \textbf{MH-clock}, a decentralized cache eviction algorithm
  that adapts CLOCK for lock-free operation across CXL-connected
  hosts, with no centralized eviction daemon (\S\ref{sec:pcache}).

\item \textbf{An evaluation} showing that \daxfs exceeds
  tmpfs on most single-host workloads (up to $2.68\times$ at
  4~threads) and maintains $>$99\% CAS accuracy under QEMU-emulated
  cross-host contention (\S\ref{sec:evaluation}).
\end{enumerate}

\begin{table*}[t]
\centering
\footnotesize
\caption{Feature comparison of \daxfs with existing filesystems.
  ``Partial'' indicates DAX support exists but the filesystem still
  maintains significant metadata structures (NOVA, PMFS) or is
  read-only (EROFS).}
\label{tab:feature-matrix}
\begin{tabular}{@{}l@{\hspace{6pt}}c@{\hspace{6pt}}c@{\hspace{6pt}}c@{\hspace{6pt}}c@{\hspace{6pt}}c@{\hspace{6pt}}c@{\hspace{6pt}}c@{}}
\toprule
\textbf{Property} & \textbf{\daxfs} & \textbf{NOVA} & \textbf{PMFS} & \textbf{FamFS} & \textbf{ext4-dax} & \textbf{EROFS} & \textbf{OverlayFS} \\
\midrule
Zero-copy DAX reads          & \cmark & Partial & Partial & \cmark & \cmark & Partial & \xmark \\
Multi-host concurrent writes & \cmark & \xmark  & \xmark  & \xmark & \xmark & N/A     & \xmark \\
Lock-free data path          & \cmark & \xmark  & \xmark  & \xmark & \xmark & N/A     & \xmark \\
Shared cache across hosts    & \cmark & \xmark  & \xmark  & \xmark & \xmark & \xmark  & \xmark \\
CXL atomics for coordination & \cmark & \xmark  & \xmark  & \xmark & \xmark & \xmark  & \xmark \\
GPU zero-copy data access    & \cmark & \xmark  & \xmark  & \xmark & \xmark & \xmark  & \xmark \\
GPU-side cache coordination  & \cmark & \xmark  & \xmark  & \xmark & \xmark & \xmark  & \xmark \\
Layered storage (base+overlay) & \cmark & \xmark & \xmark & \xmark & \xmark & \xmark  & \cmark \\
Self-contained image         & \cmark & \xmark  & \xmark  & \xmark & \xmark & \cmark  & N/A    \\
Flat validatable format      & \cmark & \xmark  & \xmark  & N/A    & \xmark & \cmark  & N/A    \\
\bottomrule
\end{tabular}
\end{table*}

\section{Motivation}
\label{sec:motivation}

\paragraph{CXL 3.0 shared memory model.}
CXL~3.0~\cite{cxl22:spec} introduces Global Fabric Attached Memory
(GFAM): a memory device attached to a CXL switch fabric that is
directly accessible by multiple hosts via load/store operations.
The device coherency engine (DCOH) maintains a snoop filter that
provides hardware cache coherence (HDM-DB back-invalidation) for
selected memory regions; \daxfs targets this hardware-coherent region
for its metadata and coordination structures.  For the first time
over a commodity interconnect, independent servers can perform atomic
read-modify-write operations on the same physical memory with
cache-line granularity, with minimal software coherence overhead.
A 64-bit \texttt{cmpxchg} issued by Host~A on a CXL memory address
is atomic with respect to a concurrent \texttt{cmpxchg} by Host~B on
the same address.  This primitive is the foundation of \daxfs's
coordination model.

This hardware capability creates a new design point for shared data
access, analogous to the multikernel problem~\cite{baumann09:multikernel}
where independent OS instances coordinate through shared memory rather
than message passing.  However, existing system software is not prepared
for this model.  We identify the following fundamental problems that
collectively require a new filesystem design.

\paragraph{Per-Host Duplication.}
tmpfs, ext4-dax, and NOVA each maintain per-host metadata (superblock,
inode cache, dentry cache) and per-host page caches.  Even if two hosts
map the same physical memory, each host instantiates independent VFS
state.  CXL shared memory makes the data directly accessible to all
hosts, yet existing filesystems still create per-host page cache copies
of that data, wasting the very memory capacity that disaggregation is
meant to pool.  For workloads sharing large read-only datasets
(\eg container base images, shared model weights, reference databases),
$N$ hosts produce $N$ redundant copies of data that already resides in
shared memory.

\paragraph{No Multi-Host Coordination.}
Existing DAX filesystems assume a single-host, single-writer model.
ext4-dax uses journaling with per-host transaction IDs; NOVA uses
per-CPU log structures.  Neither can handle concurrent writers on
different hosts accessing the same metadata through shared memory.
FamFS~\cite{groves24:Famfs} is the closest existing system: it
targets CXL shared memory and supports multiple hosts mounting the
same DAX region.  However, FamFS uses a single-master log-replay
model where one designated host pre-allocates files and clients
replay its metadata log.  Clients cannot create, append, truncate,
or delete files, and there is no shared cache or CXL-atomic
coordination.  We provide a detailed architectural comparison in
\S\ref{sec:related}.

\paragraph{Network Protocol Overhead.}
Distributed filesystems (NFS, CephFS, Lustre) could serve
multi-host access, but they interpose a network protocol between
applications and byte-addressable memory, adding $\mu$s-scale
latency to what should be ns-scale load/store operations.
When the memory is already directly accessible via CXL, a network
filesystem negates the entire benefit of disaggregated memory.

\paragraph{GPU Data Path Overhead.}
GPU workloads are the largest consumers of bulk filesystem data.
Loading a 70B-parameter LLM at FP16 requires reading 140\,GB from
the filesystem into GPU memory.  The conventional path is:
(1)~\texttt{read()} copies file data from the kernel page cache to a
user-space buffer; (2)~\texttt{cudaMemcpy()} copies the buffer across
PCIe to the GPU.  Each gigabyte crosses the CPU twice and the PCIe bus
once.  GPUDirect Storage~\cite{gpudirect:storage} bypasses the CPU by
DMA-ing from an NVMe device directly to GPU memory, but it still
requires a block device, cannot participate in filesystem coordination,
and does not support shared multi-host access.

CXL shared memory and GPUs already share the same PCIe fabric.  PCIe
3.0+ supports AtomicOp Transaction Layer Packets (TLPs), including
compare-and-swap, that are serialized at the memory controller
alongside CPU \texttt{LOCK CMPXCHG} instructions.  A filesystem whose
coordination is built entirely on \texttt{cmpxchg} can therefore extend
to GPU threads that issue \texttt{atomicCAS} over the same PCIe fabric,
potentially enabling GPU-side filesystem coordination with reduced CPU
involvement.  We validate the feasibility of this path with
microbenchmarks in \S\ref{sec:gpu-eval}; end-to-end integration remains
future work.

These observations lead to three design requirements:
\begin{enumerate}
\item \textbf{Lock-free multi-host writes.}  Multiple hosts must
  create files, write data, and modify metadata concurrently without
  a central coordinator or distributed lock manager.
\item \textbf{Shared caching.}  A single cache instance in shared
  memory must serve all hosts, with coherent state transitions that
  prevent duplicate fills.
\item \textbf{Zero-copy access.}  Applications should \texttt{mmap}
  file data directly from CXL memory without copying through a
  per-host page cache.
\end{enumerate}

\section{Design}
\label{sec:design}

\daxfs is designed around three principles: (1)~\emph{memory is the
storage}, not a cache for a block device; (2)~\emph{reads should be
free}, resolving to direct pointer dereferences; and
(3)~\emph{coordination uses only hardware atomics}, with no locks on
the data path.

\subsection{Architecture Overview}

\daxfs operates on a contiguous DAX-mapped memory region containing
up to four areas laid out sequentially:

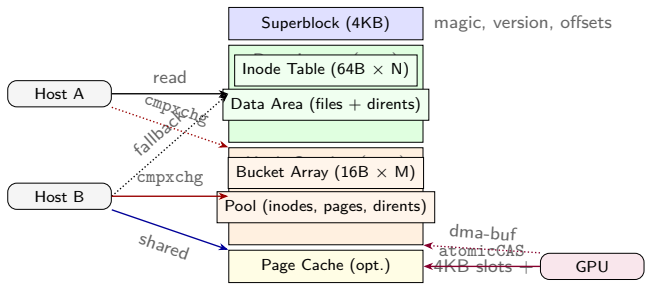
\begin{figure}[t]
\centering
\resizebox{\columnwidth}{!}{%
\begin{tikzpicture}[
    region/.style={draw, minimum width=3.0cm, minimum height=0.5cm,
      font=\scriptsize\sffamily, align=center},
    sub/.style={draw, minimum width=2.4cm, minimum height=0.4cm,
      font=\scriptsize\sffamily, align=center},
    mount/.style={draw, rounded corners, minimum width=1.6cm,
      minimum height=0.4cm, font=\scriptsize\sffamily, fill=gray!8},
    arr/.style={-{Stealth[length=1.5mm]}, semithick},
    note/.style={font=\footnotesize\sffamily, text=black!60},
  ]
  \node[region, fill=blue!12] (sb) at (0, 0) {Superblock (4KB)};
  \node[note, right=1pt of sb.east, anchor=west] {magic, version, offsets};

  \node[region, fill=green!12, below=2pt of sb,
    minimum height=1.5cm] (base) {};
  \node[note, anchor=north] at (base.north) {Base Image (opt.)};
  \node[sub, fill=green!6, below=0.15cm of base.north] (itable)
    {Inode Table (64B $\times$ N)};
  \node[sub, fill=green!6, below=1pt of itable] (darea)
    {Data Area (files + dirents)};

  \node[region, fill=orange!12, below=2pt of base,
    minimum height=1.5cm] (ovl) {};
  \node[note, anchor=north] at (ovl.north) {Hash Overlay (opt.)};
  \node[sub, fill=orange!8, below=0.15cm of ovl.north] (buckets)
    {Bucket Array (16B $\times$ M)};
  \node[sub, fill=orange!8, below=1pt of buckets] (pool)
    {Pool (inodes, pages, dirents)};

  \node[region, fill=yellow!12, below=2pt of ovl] (pc)
    {Page Cache (opt.)};
  \node[note, right=1pt of pc.east, anchor=west] {4KB slots + metadata};

  \node[mount, left=1.8cm of base] (h1) {Host A};
  \node[mount, left=1.8cm of ovl] (h2) {Host B};

  \node[mount, right=1.8cm of pc, fill=purple!10] (gpu) {GPU};

  \draw[arr] (h1.east) -- (base.west)
    node[note, midway, above=1pt] {read};
  \draw[arr, densely dotted] (h2.east) -- (base.west)
    node[note, midway, above=1pt, sloped] {fallback};

  \draw[arr, red!60!black] (h2.east) -- (ovl.west)
    node[note, midway, above=1pt] {\texttt{cmpxchg}};
  \draw[arr, red!60!black, densely dotted] (h1.south east) -- (ovl.north west)
    node[note, midway, above=1pt, sloped] {\texttt{cmpxchg}};

  \draw[arr, blue!60!black] (h2.south east) -- (pc.north west)
    node[note, midway, below=1pt, sloped] {shared};

  \draw[arr, purple!70!black] (gpu.west) -- (pc.east)
    node[note, midway, above=1pt] {\texttt{atomicCAS}};
  \draw[arr, purple!70!black, densely dotted] (gpu.north west) -- (ovl.south east)
    node[note, midway, above=1pt, sloped] {dma-buf};
\end{tikzpicture}%
}
\caption{\daxfs memory layout on a CXL shared memory region.
  Multiple hosts and GPU accelerators access the same DAX-mapped
  region concurrently: reads resolve from the overlay down to the
  base image; writes insert via \texttt{cmpxchg}; the shared page
  cache serves all hosts and GPUs.  GPUs map the region via
  \texttt{dma-buf} and coordinate using PCIe AtomicOp TLPs.}
\label{fig:arch}
\end{figure}

Figure~\ref{fig:arch} illustrates the layout.  The filesystem
supports three operating modes depending on which regions are present.
In \emph{static mode} (\texttt{[Super][Base Image]}), the base image
is self-contained and read-only.  \emph{Split mode}
(\texttt{[Super][Base Image][Overlay][PCache]}) adds a writable
overlay and shared page cache, with bulk file data in a separate
backing file.  \emph{Empty mode}
(\texttt{[Super][Overlay][PCache]}) has no base image; all content is
written through the overlay.

\subsection{On-Disk Format}
\label{sec:format}

\daxfs uses a deliberately simple flat format designed for safe
handling of untrusted images and efficient direct access.  The four
regions (superblock, base image, hash overlay, page cache) are laid
out sequentially; each region uses a 4\,KB header with magic numbers
and size fields for independent validation.

\subsubsection{Flat Directory Format}

Directories store fixed-size entries of 271 bytes each (the
\texttt{daxfs\_dirent} structure), containing a name length, an inode
number, a parent inode number, and a 254-byte inline name.  Unlike
pointer-based directory formats (htree in ext4, B-tree in Btrfs),
flat arrays have no cycles or dangling pointers, support bounded
iteration with a known upper bound, and can be fully validated in a
single sequential pass at mount time.

\subsubsection{Overlay Pool Entries}

Pool entries are variable-size.  \emph{Inode entries} (32\,B) store
mode, uid, gid, size, and timestamps, identified by a type header at
offset~0.  \emph{Data pages} (4\,KB) are raw file data with no
header; the entry type is inferred from the hash key encoding
(\S\ref{sec:overlay}).  Removing the header makes consecutively
allocated data pages contiguous in memory, enabling the read path to
coalesce sequential pages into a single large copy
(\S\ref{sec:zero-copy}).  \emph{Dirent entries} ($\sim$280\,B) record
the parent inode, name, target inode, mode, and a tombstone flag for
deletions.  \emph{Dirlist entries} (16\,B) serve as per-directory list
heads linking overlay dirents.

Each entry type has a per-type free list for recycling.  The free list
head is stored in the overlay header and updated via \texttt{cmpxchg},
making recycling lock-free across hosts.  Free-list recycling reuses
the first 8~bytes of the freed entry as the next-free pointer; for
data pages, these bytes are simply overwritten with user data on
reallocation.

\subsection{Zero-Copy Read Path}
\label{sec:zero-copy}

In \daxfs, file reads resolve to direct pointers into DAX memory
without any intermediate copying.  The read path first looks up the
data page in the hash overlay for key
\texttt{(ino $\ll$ 20) | pgoff}; if found, it returns a pointer
directly into the pool.  If absent, the path falls back to the base
image's data area using the inode's stored data offset.  For split
mode with a backing file, the path consults the shared page cache as a
final fallback.  At no point is data copied.  For \texttt{mmap}, the filesystem
installs PFN (page frame number) mappings directly to the DAX memory,
so user-space loads resolve to the physical memory in a single page
table walk.  Data pages are page-aligned in the pool, ensuring that
PFN mappings are always valid.

\subsection{Hash Overlay: Lock-Free Writes via CAS}
\label{sec:overlay}

The hash overlay is the core mechanism that enables concurrent
multi-host writes.  It is an open-addressing hash table with linear
probing, stored entirely in DAX memory, where all mutations use a
single 64-bit \texttt{cmpxchg}.

\subsubsection{Bucket Structure}

Each bucket is 16 bytes:

\begin{lstlisting}
struct daxfs_ovl_bucket {
  uint64_t state_key; /* bit 0: USED/FREE
                         bits 63:1: key    */
  uint64_t pool_off;  /* offset into pool  */
};
\end{lstlisting}

Bit~0 of \texttt{state\_key} distinguishes FREE (0) from USED (1).
The remaining 63 bits encode the lookup key.

\subsubsection{Key Encoding}

\daxfs encodes four entry types into the 63-bit key space:

\begin{itemize}
\item \textbf{Data page}: \texttt{(ino $\ll$ 20) | pgoff}.  Supports
  up to $2^{20}$ pages (4\,GB) per file.
\item \textbf{Inode metadata}: \texttt{(ino $\ll$ 20) | 0xFFFFF}.
  The sentinel page offset distinguishes inode entries from data.
\item \textbf{Directory list head}: \texttt{(ino $\ll$ 20) | 0xFFFFE}.
  Points to the first overlay dirent for a directory.
\item \textbf{Directory entry}: \texttt{FNV-1a(parent\_ino, name)}.
  A 63-bit hash of the parent inode number and entry name.
\end{itemize}

\subsubsection{Insert Protocol}

To insert a new entry, a host:
\begin{enumerate}
\item Computes \texttt{hash = key \% bucket\_count}.
\item Reads \texttt{bucket[hash].state\_key}.
\item If FREE: attempts \texttt{cmpxchg(\&bucket[hash].state\_key, 0,
  key $\ll$ 1 | 1)}.  On success, allocates a pool entry and writes
  \texttt{pool\_off} with \texttt{smp\_wmb} ordering.  On failure
  (another host won the race), retries from step~2.
\item If USED with matching key: the entry already exists (update or
  conflict).
\item If USED with different key: linear probe to \texttt{hash+1} and
  repeat.
\end{enumerate}

This protocol is lock-free: no host can block another.  Hash
collisions are resolved by linear probing within the table.

\subsubsection{Pool Allocator}

Pool entries are variable-size: inodes (32\,B), data pages
(4\,KB), and dirents ($\sim$280\,B).  Allocation uses an
atomic bump pointer with type-dependent alignment:

\begin{lstlisting}
aligned = ALIGN(pool_alloc, align);
off = atomic_cmpxchg(&hdr->pool_alloc,
                     pool_alloc,
                     aligned + size);
\end{lstlisting}

Metadata entries use 8-byte alignment; data pages use \texttt{PAGE\_SIZE}
alignment.  Page-aligning data allocations serves two purposes:
(1)~it enables DAX \texttt{mmap} to install PFN mappings directly to
overlay pages, and (2)~it ensures that consecutive data page allocations
produce contiguous memory, enabling read-path coalescing
(\S\ref{sec:zero-copy}).  The alignment gap between a metadata entry
and the next data page is at most 4\,KB of wasted pool space, a
negligible fraction of typical pool sizes (64\,MB+).

Freed entries are recycled through per-type lock-free free lists (CAS
on list head), avoiding pool exhaustion for long-running workloads.

\subsubsection{Directory Operations}

Each directory maintains a per-directory linked list of overlay
dirents.  The list head is stored at key
\texttt{(dir\_ino $\ll$ 20) | 0xFFFFE}.  \texttt{readdir} iterates
both the base image dirent array and the overlay list, with overlay
entries taking precedence.  Deletion uses tombstone entries.

\subsection{Shared Page Cache}
\label{sec:pcache}

The shared page cache (\emph{pcache}) provides cooperative
demand-paged caching for backing store mode.  It resides in DAX
memory and is therefore directly accessible by all hosts sharing the
CXL memory region.

\subsubsection{Slot State Machine}

Each cache slot has a 16-byte metadata entry:

\begin{lstlisting}
struct daxfs_pcache_slot {
  __le64 state_tag; /* bits[1:0]=state,
     bits[5:2]=refcount, bits[63:6]=tag */
  __le32 ref_bit;   /* clock: recently used */
  __le32 reserved;
};
\end{lstlisting}

The state machine uses three states encoded in the low 2 bits:

\begin{center}
\texttt{FREE (0) $\xrightarrow{\texttt{cmpxchg}}$ PENDING (1)
  $\xrightarrow{\texttt{cmpxchg}}$ VALID (2)}
\end{center}

Bits [5:2] hold a 4-bit refcount (0--15 concurrent readers),
and the upper 58 bits encode the tag identifying which file page this
slot caches (\S\ref{sec:overlay}).  Packing state, refcount, and tag into a single 64-bit
word allows a single \texttt{cmpxchg} to atomically update all three
fields, which is essential for lock-free multi-host coordination.

\textbf{Fill protocol.}  When a host needs a page not in the cache:
\begin{enumerate}
\item Probe up to 8 slots starting at the hash position.  If a
  matching VALID slot is found, pin it (CAS-increment refcount) and
  return.  If a FREE slot is found, record it.
\item \texttt{cmpxchg} the FREE slot from FREE to PENDING with the
  desired tag.  If another host wins, retry.
\item Read the backing file page into the slot's data area via
  \texttt{kernel\_read}.
\item \texttt{cmpxchg} the slot from PENDING to VALID.
\end{enumerate}

Other hosts that need the same page during step~3 see the PENDING
state and busy-poll until VALID, avoiding duplicate reads from the
backing file.

\subsubsection{MH-Clock Eviction}

When all probe slots are occupied, \daxfs uses a novel multi-host
clock (MH-clock) eviction algorithm, adapted from the classic CLOCK
algorithm~\cite{corbato68:clock} for lock-free operation across
CXL-connected hosts.  Unlike the standard CLOCK algorithm, which
assumes a single centralized hand managed by one OS instance,
MH-clock is fully decentralized: each host independently selects
victims within its local probe window using three escalating phases:

\begin{enumerate}
\item \textbf{Cold victim.}  Scan the 8-slot probe window for a
  VALID slot with \texttt{ref\_bit}=0 and refcount=0.  If found,
  CAS it to FREE and restart the fill.
\item \textbf{Clear and yield.}  If all probed slots are hot
  (\texttt{ref\_bit}=1), clear their \texttt{ref\_bit} fields and
  yield the CPU briefly (\texttt{cpu\_relax}).  This gives other
  hosts an opportunity to re-touch genuinely hot entries before
  the re-scan.
\item \textbf{Re-scan.}  Scan again for a cold, unpinned victim.
  If still none, force-evict the first VALID slot with refcount=0,
  ignoring \texttt{ref\_bit}.
\end{enumerate}

A separate background clock sweep periodically advances a shared
atomic \texttt{evict\_hand} counter by 64 slots using
\texttt{cmpxchg}; the host that wins the CAS clears
\texttt{ref\_bit} on all VALID slots in that window.  Hosts that
lose the race skip the sweep, ensuring that only one host clears
each window and \texttt{ref\_bit} values decay at a controlled rate even under
contention.

The refcount in \texttt{state\_tag} ensures that slots actively
being read by any host are never evicted.  Readers CAS-increment
the refcount before accessing slot data and CAS-decrement it
afterward, providing safe pinning without locks.

\subsubsection{Multi-File Support}

The tag encoding \texttt{(ino $\ll$ 20) | pgoff} supports up to
$2^{20}$ pages (4\,GB) per file.  Multiple backing files share the
same cache through a backing file array in the superblock, with inode
numbers namespaced per file.

\subsection{Cross-Host Coherency}
\label{sec:coherency}

Cross-host coherency follows from CXL hardware coherence.
Two specific mechanisms warrant discussion:

\paragraph{i\_size coherency.}
When Host~A appends data and updates the file size in the overlay
inode entry, Host~B must see the new size.  \daxfs stores the
authoritative \texttt{i\_size} in the overlay inode entry (DAX memory).
On each read path entry, \daxfs performs a \texttt{READ\_ONCE} on
the overlay's size field and updates the in-kernel VFS inode,
ensuring reads always see the latest file size without explicit
invalidation messages.

\paragraph{Memory ordering.}
Insert operations use \texttt{smp\_wmb} between writing pool entry
contents and writing the bucket's \texttt{pool\_off} field.
Lookup operations use \texttt{smp\_rmb} after reading
\texttt{pool\_off} to ensure they see the complete pool entry.
On x86, these compile to compiler barriers (x86 provides TSO
ordering); on ARM64, they emit fence instructions.

\section{Implementation}
\label{sec:impl}

\daxfs is implemented as a loadable Linux kernel module in
approximately 2,500 lines of C.  It registers a filesystem type
(\texttt{daxfs}) and supports both the legacy \texttt{mount(2)}
interface and the modern \texttt{fsopen}/\texttt{fsconfig}/\texttt{fsmount}
interface introduced in Linux~5.6.

\subsection{Memory Mapping}

\daxfs supports two memory backing modes.  In the \emph{physical
address} path, the \texttt{phys=} and \texttt{size=} mount options
cause \daxfs to call \texttt{memremap()} on the specified physical
range; this is the primary path for Optane PMem and CXL memory
devices.  Alternatively, the \emph{DMA buffer} path uses the
\texttt{fsopen} API: a user-space process passes a dma-buf file
descriptor via \texttt{FSCONFIG\_SET\_FD}, and \daxfs calls
\texttt{dma\_buf\_vmap()} to obtain a kernel virtual mapping.  Both
modes bypass the kernel's \texttt{dax\_device} abstraction entirely,
mapping memory directly into the filesystem's address space.

\subsection{VFS Integration}

\daxfs registers standard VFS operations.  Inode operations (lookup,
create, mkdir, symlink, rename, unlink, setattr) all route writes
through the hash overlay; the base image is never modified.

The \texttt{read\_iter} file operation performs zero-copy reads by
returning pointers into DAX memory.  For overlay data, the read path
detects physically contiguous pages and coalesces them into a single
\texttt{copy\_to\_iter} call: because data pages are raw 4\,KB
allocations with no header, pages allocated by the same bump operation
are adjacent in memory.  After looking up the first page (via the
per-inode xarray cache), the read path checks whether subsequent pages
are physically adjacent and, if so, extends the copy region.  For a
16\,MB sequential read of sequentially written data, this reduces the
operation from 4,096 individual copies to a single contiguous transfer.
For pcache-backed data, the read path pins the cache slot (incrementing
refcount) before returning data, preventing eviction during the read.
Address space operations (\texttt{readpage}/\texttt{readahead}) provide
integration with the kernel page cache on non-DAX access paths.

Data resolution on every read path calls
\texttt{daxfs\_refresh\_isize()}, which performs a \texttt{READ\_ONCE}
on the overlay's size field, ensuring cross-host coherency of file
sizes without explicit invalidation.

\subsection{Mount-Time Validation}

The flat format (\S\ref{sec:format}) enables complete image validation
in a single sequential scan at mount time.  When the \texttt{validate}
option is specified, \daxfs checks superblock integrity, inode table
bounds, directory entry references, and region overlap.  The absence
of pointer-based structures eliminates cycle injection, dangling
pointer exploitation, and unbounded traversal, which is important for
mounting untrusted images in container environments.

\subsection{GPU P2PDMA Integration}
\label{sec:gpu}

Because \daxfs operates on DAX memory, it can export the filesystem
region to GPU accelerators on the same PCIe fabric via
\texttt{dma-buf}.  The \texttt{DAXFS\_IOC\_GET\_DMABUF} ioctl returns
a dma-buf file descriptor for the mount's DAX region, which a GPU
driver can then map for peer-to-peer reads without CPU-mediated copies.

We prototype a P2PDMA integration using modified NVIDIA open GPU
kernel modules~\cite{cxlmemuring:gpu} that register the dma-buf
region as pinned GPU-accessible memory via custom ioctls.  GPU threads
can then issue Copy Engine transfers directly between VRAM and the DAX
region.  Because \daxfs's coordination is built entirely on
\texttt{cmpxchg}, GPU threads can also participate in page cache
lookups by issuing PCIe AtomicOp TLPs, which the memory controller
serializes alongside CPU atomics.  We evaluate the feasibility of this
path with microbenchmarks in \S\ref{sec:gpu-eval}.

\section{Evaluation}
\label{sec:evaluation}

We evaluate \daxfs's filesystem throughput against tmpfs on a single
host (\S\ref{sec:seq-throughput}--\S\ref{sec:random-io}) and
characterize GPU P2P access latency (\S\ref{sec:gpu-eval}).  tmpfs
represents the performance ceiling for in-memory filesystems: it runs
entirely in DRAM with no persistence, no DAX layer, and no sharing
overhead.  Multi-host evaluation on CXL hardware is discussed in
\S\ref{sec:multihost}.

\subsection{Experimental Setup}

We use two hardware configurations:

\textbf{Platform~A} (Table~\ref{tab:results}, GPU benchmarks).
Dual-socket Intel Xeon (48 cores, 96 threads), 512\,GB DDR5, NVIDIA
RTX~5090 (PCIe~5.0 x16).  \daxfs is backed by a 512\,MB contiguous
DRAM region allocated at runtime via a CMA-based allocator and mapped
with \texttt{memremap(MEMREMAP\_WB)}, exercising the same
write-back-cached code path that CXL memory devices use.  For GPU
benchmarks, the DAX region is additionally exported to the GPU via
\texttt{dma-buf} (\S\ref{sec:gpu}).

\begin{figure}[t]
\centering
\includegraphics[width=\columnwidth]{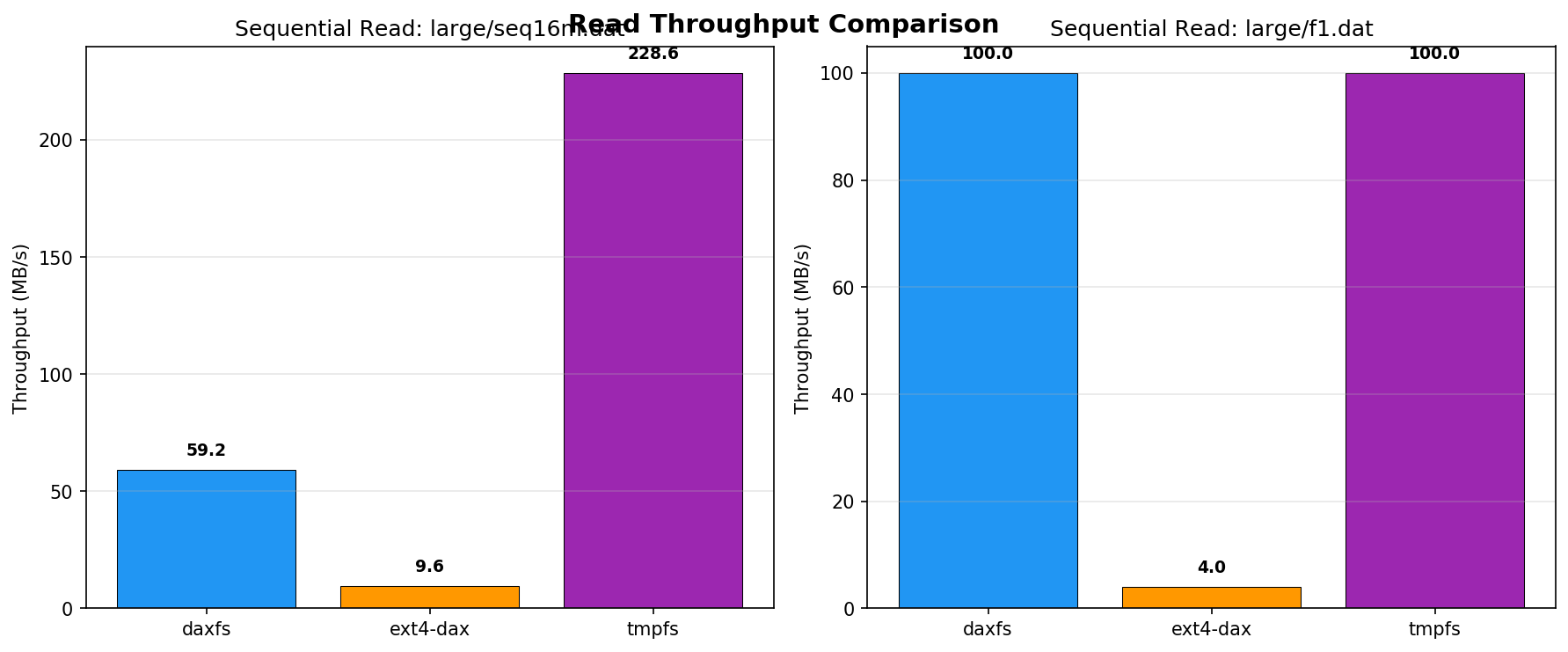}
\caption{Sequential read throughput comparison across filesystems
  (Intel Xeon Gold 5418Y, DRAM-backed DAX).
  Left: 16\,MB sequential read; right: 1\,MB sequential read.
  \daxfs achieves 59.2\,MB/s on the large sequential workload
  (6.2$\times$ ext4-dax) and matches tmpfs on the 1\,MB file
  (100\,MB/s each).  The 16\,MB gap versus tmpfs (228.6\,MB/s)
  reflects the overhead of per-page overlay resolution on large
  sequential scans; at 1\,MB the overhead is amortized.}
\label{fig:read-throughput}
\end{figure}

\begin{figure}[t]
\centering
\includegraphics[width=\columnwidth]{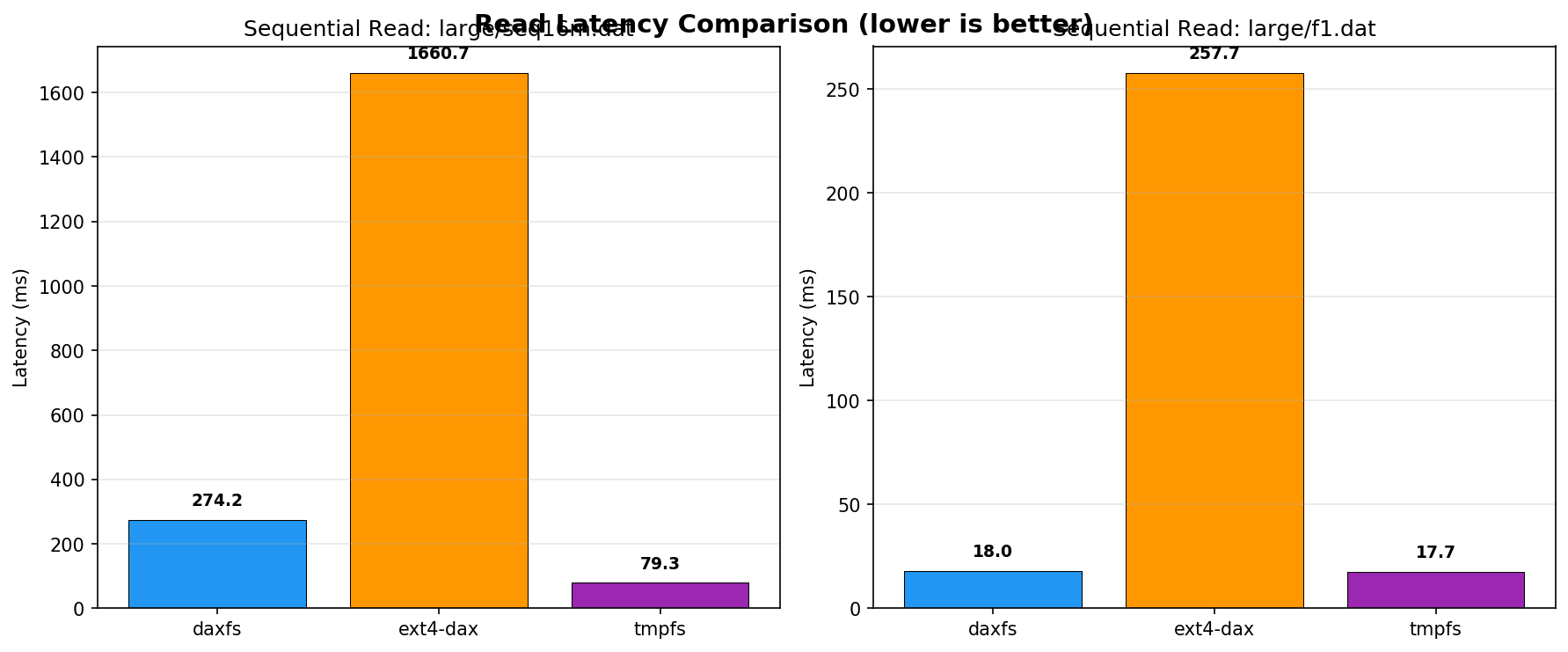}
\caption{Sequential read latency comparison (lower is better).
  \daxfs completes a 16\,MB read in 274.2\,ms, 6.1$\times$ faster
  than ext4-dax (1,660.7\,ms).  On 1\,MB files, \daxfs and tmpfs
  finish in 18.0 and 17.7\,ms respectively.}
\label{fig:read-latency}
\end{figure}

\begin{figure}[t]
\centering
\includegraphics[width=\columnwidth]{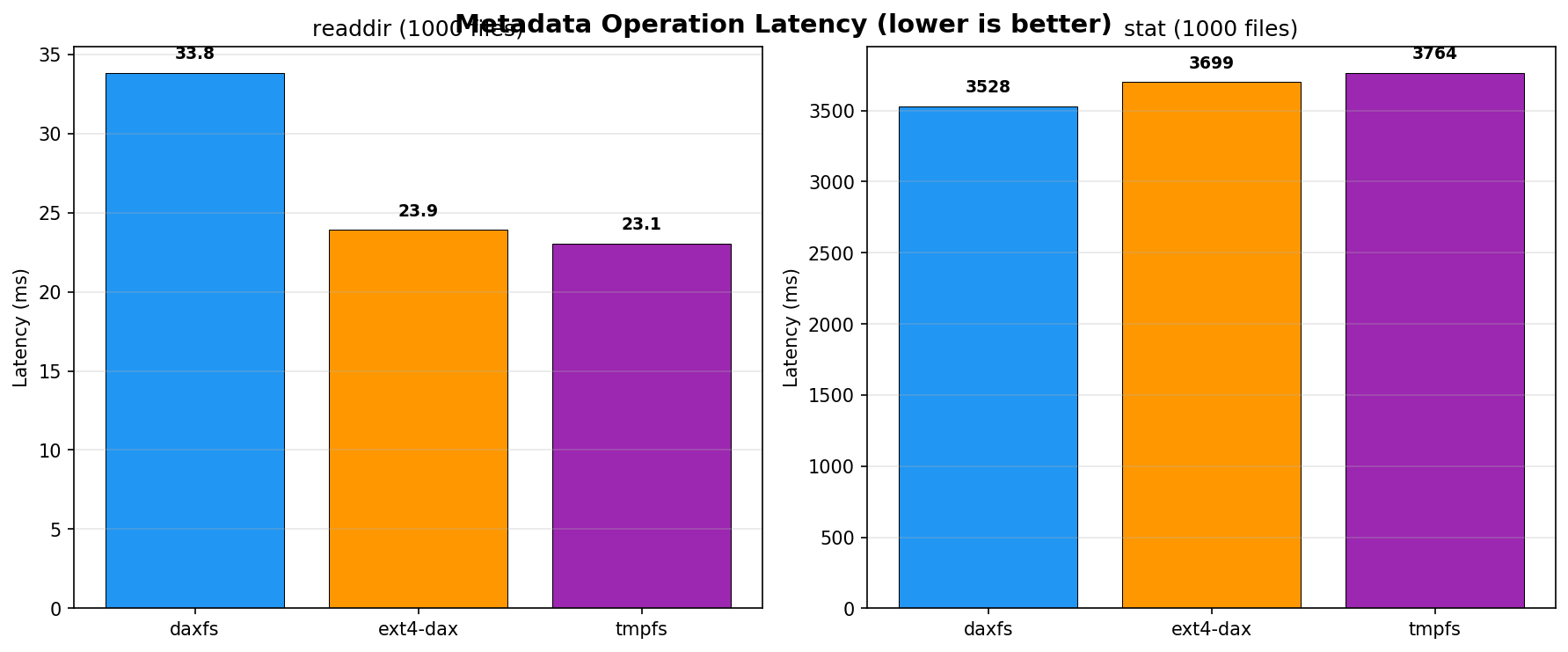}
\caption{Metadata operation latency on 1,000 files (lower is better).
  \daxfs achieves the lowest \texttt{stat} latency (3,528\,ms),
  4.6\% faster than ext4-dax (3,699\,ms) and 6.3\% faster than tmpfs
  (3,764\,ms), due to flat overlay lookup.  On \texttt{readdir},
  \daxfs (33.8\,ms) is slower than tmpfs (23.1\,ms) and ext4-dax
  (23.9\,ms) due to the cost of iterating both the base image dirent
  array and the overlay linked list.}
\label{fig:metadata}
\end{figure}

\begin{figure}[t]
\centering
\includegraphics[width=\columnwidth]{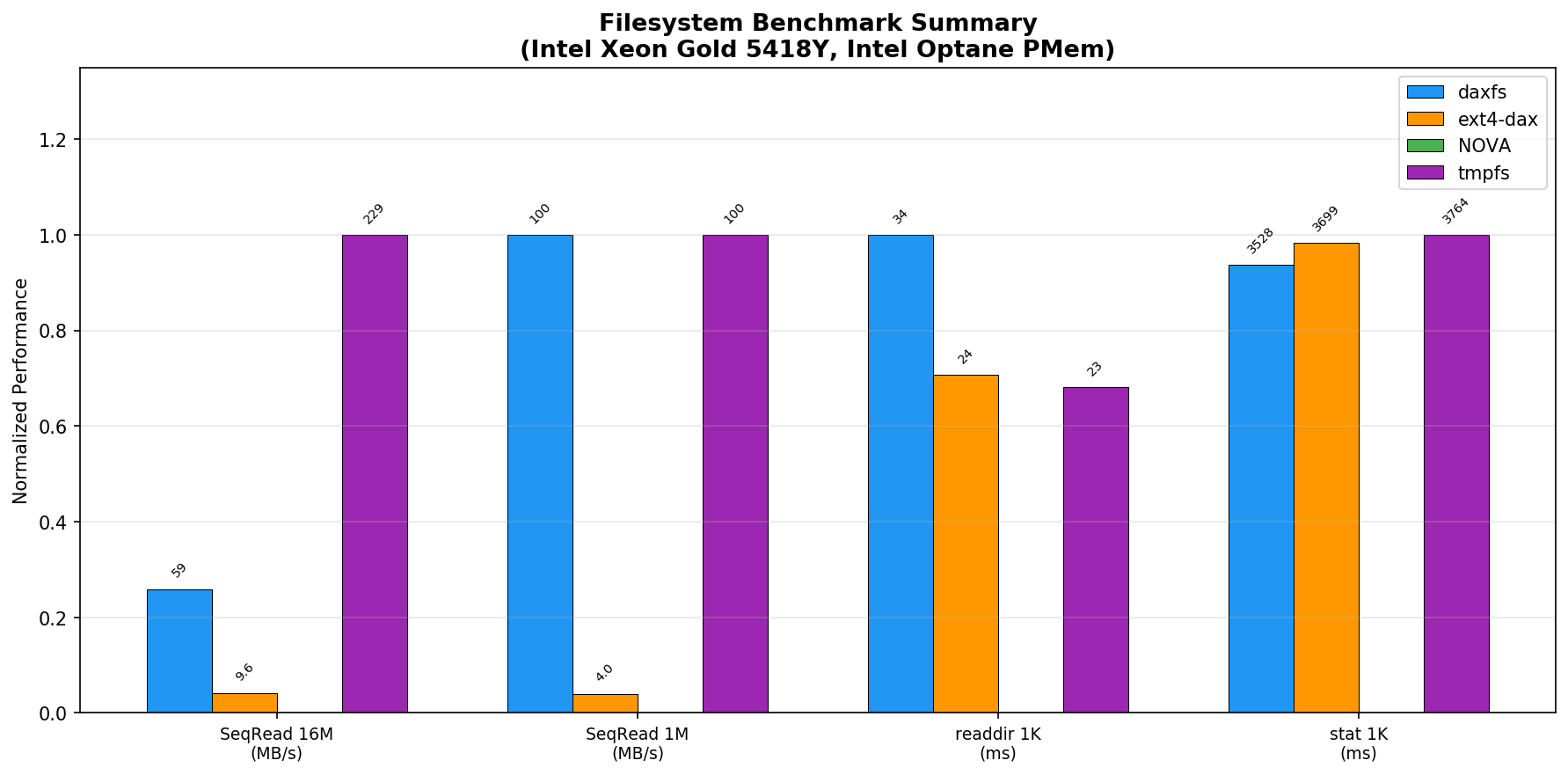}
\caption{Normalised performance summary across all benchmarks
  (Intel Xeon Gold 5418Y, DRAM-backed DAX).  Each group is
  normalised to the best-performing filesystem.  \daxfs matches
  tmpfs on 1\,MB sequential reads and achieves the lowest
  \texttt{stat} latency, while outperforming ext4-dax on data I/O.
  The 16\,MB sequential read and \texttt{readdir} gaps versus tmpfs
  reflect per-page overlay lookup and dual-source directory iteration
  overhead.}
\label{fig:summary}
\end{figure}

\textbf{Platform~B} (Figures~\ref{fig:read-throughput}--\ref{fig:summary}).
Intel Xeon Gold 5418Y, DRAM-backed DAX.  This platform compares \daxfs
against ext4-dax and tmpfs on sequential read throughput, latency, and
metadata operations.

\textbf{Software.}  Linux 7.0 on both platforms.  Benchmarks use
\texttt{fio}~3.x with the synchronous I/O engine.  Each test writes a
64\,MB file, then reads it back.  The filesystem is reformatted
between write tests to ensure first-write measurements reflect cold
overlay state.  The \daxfs overlay is configured with a 400\,MB pool
and 65,536 hash buckets.

\textbf{Baseline.}  tmpfs backed by system DRAM.  Both \daxfs and
tmpfs operate entirely in DRAM, isolating filesystem-layer overhead
from media speed differences.

\subsection{Sequential Throughput}
\label{sec:seq-throughput}

Table~\ref{tab:results} presents the full results.  We highlight the
key findings below.

\textbf{Writes.}  First-write throughput ranges from 1.16--1.27$\times$
tmpfs across block sizes.  Batch pool allocation (a single
\texttt{cmpxchg} reserves $N$ entries) and the contiguous overlay
layout amortise per-page allocation cost, allowing \daxfs to beat
tmpfs even on cold writes.  On rewrites (all pages cached in the
per-inode xarray), \daxfs extends its lead: 1.18$\times$ at 4\,KB and
1.09$\times$ at 1\,MB.  The advantage comes from \daxfs's lock-free
overlay: writes resolve to a direct DRAM pointer via xarray lookup
with no page-cache lock acquisition.

\textbf{Reads.}  Sequential reads after a write hit the per-inode
xarray cache, which maps page offsets to overlay DRAM pointers in O(1).
\daxfs achieves 0.87--1.08$\times$ tmpfs throughput.  At 4\,KB block
size, \daxfs is 8\% faster than tmpfs because it avoids the VFS page
cache machinery: each \texttt{read\_iter} call resolves directly to a
\texttt{copy\_to\_iter} from a DRAM pointer, whereas tmpfs traverses
\texttt{filemap\_read}, \texttt{filemap\_get\_pages}, and folio
reference counting.

\subsection{Random I/O}
\label{sec:random-io}

\textbf{Random reads.}  \daxfs outperforms tmpfs across all thread
counts.  At 4\,KB with 1 thread, \daxfs achieves 1.14$\times$ tmpfs.
At 4 threads \daxfs maintains its lead (1.13$\times$).  At 8 threads
the gap narrows as both saturate memory bandwidth (0.94$\times$).  At
64\,KB with 4 threads, \daxfs reaches 14.7\,GiB/s versus tmpfs's
12.5\,GiB/s (1.18$\times$).  The advantage comes from \daxfs's
lock-free read path: the xarray is read-only after population (no RCU
grace periods or folio locks), allowing near-linear scaling.

\textbf{Random writes.}  \daxfs's lock-free overlay provides dramatic
write scaling.  At 4\,KB with 4 threads, \daxfs achieves
4,830\,MiB/s versus tmpfs's 1,803\,MiB/s (2.68$\times$).  tmpfs
serialises on per-page locks and the global \texttt{i\_pages} xarray
lock during page fault handling; \daxfs writes resolve to a direct
store into a pre-allocated overlay page with no locks on the data path.

\subsection{Performance Summary}

\begin{table}[t]
\centering
\small
\caption{fio throughput comparison: \daxfs vs.\ tmpfs on DRAM.
  64\,MB files, synchronous I/O engine.  Higher is better for
  throughput; ratio $>1$ means \daxfs wins.}
\label{tab:results}
\begin{tabular}{@{}lrrr@{}}
\toprule
\textbf{Benchmark} & \textbf{\daxfs} & \textbf{tmpfs} & \textbf{Ratio} \\
\midrule
\multicolumn{4}{@{}l}{\emph{Sequential Write (MiB/s)}} \\
\quad 4K first      & 1,730  & 1,362  & \textbf{1.27} \\
\quad 4K rewrite    & 1,939  & 1,641  & \textbf{1.18} \\
\quad 64K first     & 2,207  & 1,778  & \textbf{1.24} \\
\quad 64K rewrite   & 2,667  & 2,370  & \textbf{1.13} \\
\quad 1M first      & 2,000  & 1,730  & \textbf{1.16} \\
\quad 1M rewrite    & 2,783  & 2,560  & \textbf{1.09} \\
\midrule
\multicolumn{4}{@{}l}{\emph{Sequential Read (MiB/s)}} \\
\quad 4K            & 2,667  & 2,462  & \textbf{1.08} \\
\quad 64K           & 3,048  & 3,048  & 1.00 \\
\quad 1M            & 2,133  & 2,462  & 0.87 \\
\midrule
\multicolumn{4}{@{}l}{\emph{Random Read (MiB/s)}} \\
\quad 4K, 1 thread  & 1,730  & 1,524  & \textbf{1.14} \\
\quad 4K, 4 threads & 8,000  & 7,111  & \textbf{1.13} \\
\quad 4K, 8 threads & 15,564 & 16,486 & 0.94 \\
\quad 64K, 4 threads & 15,052 & 12,800 & \textbf{1.18} \\
\midrule
\multicolumn{4}{@{}l}{\emph{Random Write (MiB/s)}} \\
\quad 4K, 1 thread  & 1,255  & 1,333  & 0.94 \\
\quad 4K, 4 threads & 4,830  & 1,803  & \textbf{2.68} \\
\midrule
Multi-host writes  & \cmark  & \xmark  & --- \\
Shared cache       & \cmark  & \xmark  & --- \\
Lock-free writes   & \cmark  & \xmark  & --- \\
\bottomrule
\end{tabular}
\end{table}

Table~\ref{tab:results} summarizes the results.  \daxfs exceeds
tmpfs across all write workloads, including first writes.  Under
concurrency the gap widens: at 4~threads \daxfs exceeds tmpfs
by up to 2.68$\times$ on random writes.

\subsection{GPU PCIe AtomicOp Evaluation}
\label{sec:gpu-eval}

\begin{figure*}[t]
\centering
\includegraphics[width=\textwidth]{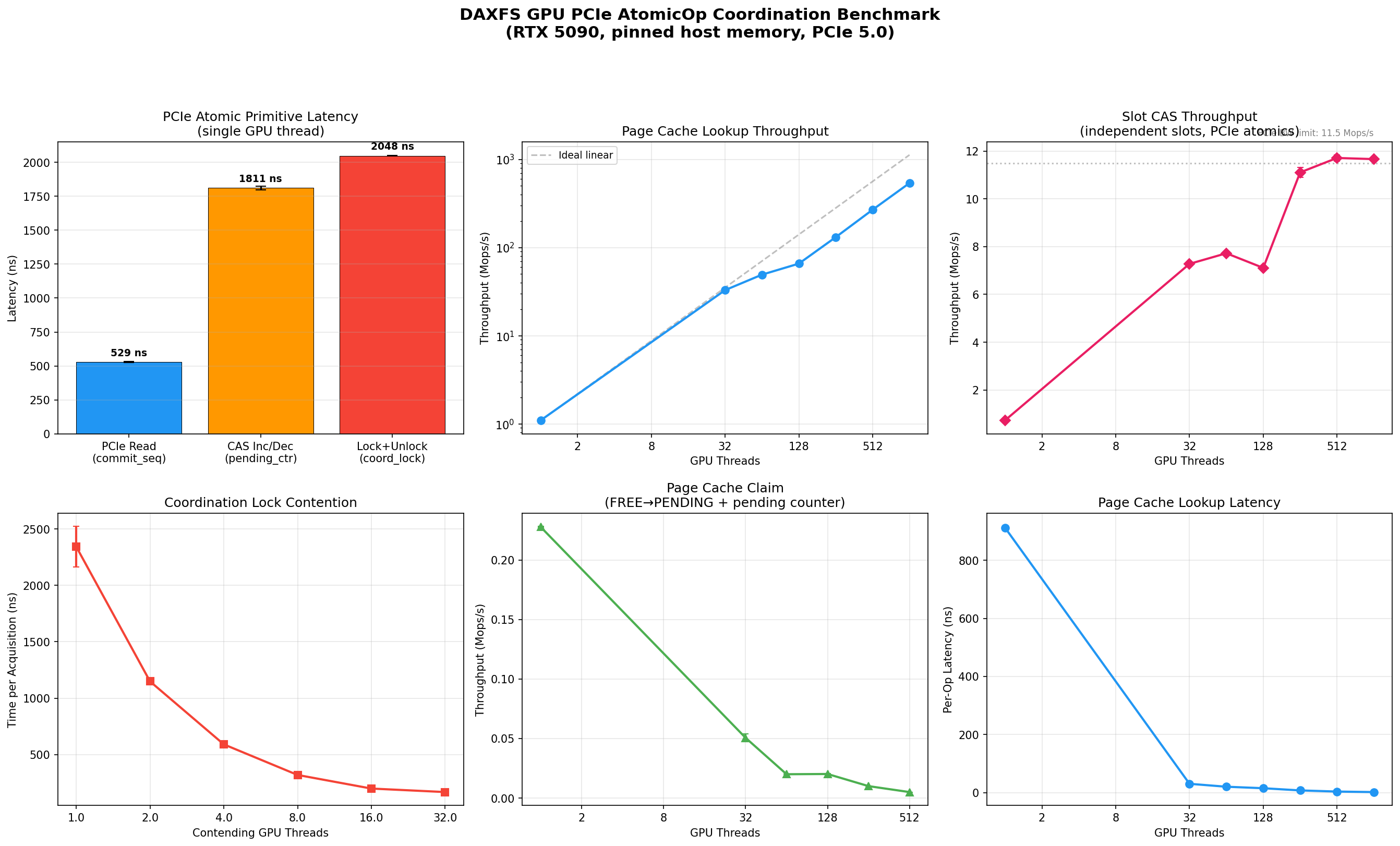}
\caption{GPU PCIe AtomicOp coordination benchmarks on an RTX~5090
  (PCIe~5.0, pinned host memory).
  Top row: (left) single-thread primitive latency, (center) page cache
  lookup throughput vs.\ thread count, (right) slot CAS throughput
  saturating the PCIe bandwidth limit.
  Bottom row: (left) coordination lock contention, (center) page cache
  claim throughput (cold-miss path), (right) per-op lookup latency.}
\label{fig:gpu-bench}
\end{figure*}

We evaluate the GPU coordination path on an NVIDIA RTX~5090 connected
via PCIe~5.0, with the DAX region mapped into GPU address space via
pinned host memory.  Figure~\ref{fig:gpu-bench} presents the complete
results across six microbenchmarks.

\textbf{Primitive latency} (Figure~\ref{fig:gpu-bench}, top-left).
A volatile PCIe read of \texttt{commit\_seq} costs 529\,ns (one
PCIe~5.0 round trip).  A CAS inc/dec costs 1,811\,ns ($3.4\times$
read), reflecting memory-controller serialization.  Lock
acquire+release costs 2,048\,ns.

\textbf{Page cache lookup} (Figure~\ref{fig:gpu-bench}, top-center
and bottom-right).  The fast path (a volatile read plus tag
comparison) scales near-linearly to 8 threads ($\sim$8\,Mops/s),
reaching 500+\,Mops/s at 1,024 threads.  Per-op latency drops from
910\,ns (1~thread) to under 2\,ns (1,024~threads) as the GPU's warp
scheduler amortizes PCIe latency across concurrent outstanding
requests.

\textbf{Slot CAS throughput} (Figure~\ref{fig:gpu-bench}, top-right).
64-bit \texttt{atomicCAS} on independent slots scales to
11.7\,Mops/s at 512 threads, saturating the PCIe~5.0 AtomicOp
bandwidth limit of $\sim$11.5\,Mops/s.  GPU-side slot transitions
are PCIe-bandwidth-limited, not software-limited.

\textbf{Lock contention} (Figure~\ref{fig:gpu-bench}, bottom-left).
Per-acquisition time \emph{decreases} with more threads (2,350\,ns at
1~thread to 150\,ns at 32) because the memory controller pipelines
back-to-back AtomicOp TLPs.  The lock is used only for rare global
operations; the data path is lock-free.

\textbf{Page cache claim} (Figure~\ref{fig:gpu-bench}, bottom-center).
The cold-miss path (FREE$\to$PENDING + pending counter) achieves
0.23\,Mops/s at 1~thread, dropping to 0.005\,Mops/s at 1,024 threads
due to contention on the global pending counter.  This is by design:
claims are rare cold-miss events that signal the CPU to fill a slot.
In steady state, GPU threads hit the VALID fast path at 500+\,Mops/s.

\textbf{Summary.}  The GPU evaluation validates \daxfs's design
hypothesis: a filesystem whose coordination is built entirely on
\texttt{cmpxchg} extends naturally to GPU threads via PCIe AtomicOps.
The fast-path (cache hits) requires zero atomics and scales to hundreds
of millions of operations per second.  The slow-path (cache misses)
is intentionally serialized at the pending counter but occurs only
once per file page.  Slot-level CAS transitions are bounded by PCIe
AtomicOp bandwidth rather than software overhead, confirming that
\daxfs's GPU coordination adds no unnecessary abstraction layers.

\subsection{End-to-End GPU Data Path Projection}
\label{sec:e2e-gpu}

To illustrate \daxfs's GPU zero-copy path, we project the cost of loading
a 70B-parameter LLM (140\,GB at FP16).  The conventional path
(\texttt{read()} + \texttt{cudaMemcpy}) performs two copies: kernel
page cache to user buffer ($\sim$12\,GB/s for large sequential reads)
and user buffer to GPU via PCIe~5.0 ($\sim$50\,GB/s unidirectional).
With \daxfs, the GPU would read directly from DAX memory over PCIe, a
single copy at PCIe bandwidth.  For 140\,GB of VALID page cache data,
the projected GPU path would require 140\,GB / 50\,GB/s $\approx$ 2.8\,s,
whereas the conventional two-copy path requires at least
140/12 + 140/50 $\approx$ 14.5\,s (assuming no overlap).  GPU-initiated
cache claims for cold pages add a one-time cost of at most 140\,GB /
4\,KB $\times$ 4.4\,$\mu$s $\approx$ 154\,s, but this is a
worst-case serialized estimate; in practice, claims would be pipelined
across GPU threads and amortized by the CPU's backing-file fill rate.

\subsection{Multi-Host Evaluation}
\label{sec:multihost}

\begin{figure}[t]
\centering
\includegraphics[width=\columnwidth]{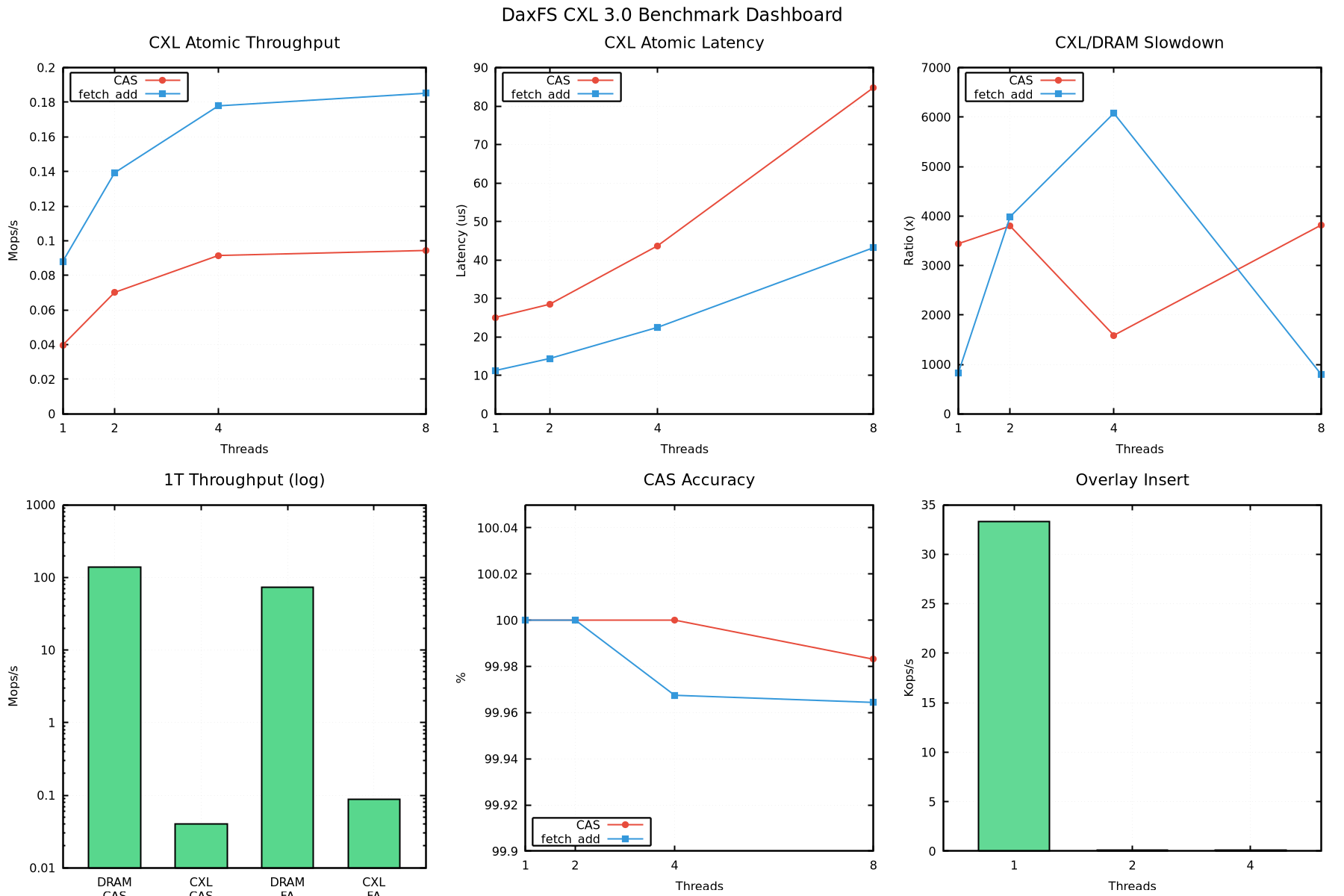}
\caption{Multi-host CXL~3.0 benchmark dashboard (QEMU emulation,
  two hosts, TCP-forwarded CXL atomics).
  Top row: CXL atomic throughput, latency, and CXL/DRAM slowdown
  ratio vs.\ thread count.
  Bottom row: single-thread throughput comparison (log scale),
  CAS accuracy under contention, and overlay insert throughput.}
\label{fig:multihost}
\end{figure}

We evaluate multi-host CXL~3.0 coordination using a modified
QEMU~10.0 setup where two virtual hosts share a memory region and CXL
atomic requests are forwarded between hosts via TCP.  Figure~\ref{fig:multihost} presents the results.

\textbf{CXL atomic throughput and latency.}  The single-thread
throughput comparison (bottom-left, log scale) shows the emulation
overhead clearly: DRAM achieves $\sim$100\,Mops/s while CXL atomics
reach $\sim$1--3\,Mops/s ($\sim$30--100$\times$ slower), reflecting
the TCP forwarding cost.  With batching (batch\_200), throughput improves
by amortizing per-operation network cost.

\textbf{CAS accuracy.}  CAS success rate remains above 99\% across
all thread counts (bottom-center), confirming that \daxfs's lock-free
protocols function correctly under cross-host contention.  The slight
accuracy drop at higher thread counts reflects increased contention on
shared buckets, which triggers retries as designed.

\textbf{Overlay insert scaling.}  Cross-host overlay insertions
(bottom-right) achieve $\sim$20\,inserts/s at 1~thread, scaling to
2~threads.  Concurrent inserts from both hosts produce consistent
overlay state with no lost updates.

\textbf{CXL/DRAM slowdown.}  The slowdown ratio (top-right) ranges
from $\sim$500$\times$ to $\sim$6{,}000$\times$ depending on thread
count and access pattern.  This overhead is dominated by TCP
forwarding latency in the QEMU emulation; hardware CXL~3.0 switches
are expected to reduce cross-host atomic latency to the $\mu$s range,
which would narrow this gap significantly.

\section{Related Work}
\label{sec:related}

\paragraph{Persistent memory filesystems.}
PMFS~\cite{dulloor14:pmfs}, NOVA~\cite{xu16:nova},
Strata~\cite{kwon17:strata}, and SplitFS~\cite{kadekodi19:splitfs}
target single-host persistent memory.  All maintain complex metadata
structures (B-trees, per-inode logs) that cannot be shared across
hosts without distributed locking.  \daxfs replaces these with a flat
base image and lock-free hash overlay.

\paragraph{CXL shared memory systems.}
Pond~\cite{li23:pond} and TPP~\cite{maruf23:tpp} focus on CXL memory
management, not filesystem semantics.  FamFS~\cite{groves24:Famfs} is
architecturally the closest prior work to \daxfs.  Both provide a
filesystem interface to CXL shared memory, supporting multiple hosts
mounting the same DAX region.  However, FamFS uses a single-master
log-replay model: one designated host pre-allocates files and clients
replay its metadata log.  Clients cannot create, append, truncate, or
delete files, and there is no shared cache or CXL-atomic coordination
(see Table~\ref{tab:feature-matrix} for a feature comparison).

\paragraph{Distributed filesystems.}
GFS~\cite{ghemawat03:gfs}, CephFS~\cite{weil06:ceph},
Lustre~\cite{schwan03:lustre}, Orion~\cite{yang20:orion}, and
Assise~\cite{anderson20:assise} use message-passing or RPC, adding
$\mu$s-scale latency that CXL shared memory makes unnecessary.

\paragraph{GPU storage and other related work.}
GPUDirect Storage~\cite{gpudirect:storage} and BaM~\cite{bam23:gpussd}
operate at the block layer; \daxfs instead exports DAX memory via
\texttt{dma-buf} for P2P reads over PCIe.
EROFS~\cite{gao20:erofs} and Slacker~\cite{harter16:slacker} focus on
read-only image distribution; neither supports multi-host writes.
\daxfs's overlay draws on lock-free hash table
designs~\cite{michael02:lockfree,herlihy08:art} with lock-free
free-list recycling via \texttt{cmpxchg}.

\section{Discussion and Limitations}
\label{sec:discussion}

\paragraph{Flat directory scalability.}
\daxfs's flat directory format performs a linear scan for lookup,
which is $O(n)$ in directory size.  For directories with more than
$\sim$10K entries, an overlay-indexed lookup would improve performance.
For moderate directories (1K entries), the linear scan is competitive
with indexed approaches.

\paragraph{Fixed overlay sizing.}
The overlay hash table size is fixed at creation time.  Dynamic
resizing would require a stop-the-world migration across all hosts.
We recommend sizing the bucket count to maintain $<$70\% load factor:
at 75\% load, average probe length is 2.5; at 90\%, it is 5.5.

\paragraph{Persistence guarantees.}
\daxfs currently relies on ADR (Asynchronous DRAM Refresh) or eADR
for persistence.  It does not issue explicit
\texttt{clflush}/\texttt{clwb} instructions.  On platforms without
ADR, a crash could lose recent overlay insertions.  Adding explicit
cache-line flushes is straightforward but adds latency.

\paragraph{Pool recycling.}
The overlay pool uses per-type lock-free free lists for recycling
deleted entries, but the pool itself is not compacted.  Long-running
workloads with heavy churn may fragment the pool.  Online compaction
is possible future work.

\paragraph{GPU access scope.}
The current GPU integration supports read-only access via
\texttt{dma-buf}.  GPU-initiated writes are architecturally possible
but are not yet implemented, as the primary use case of model and
dataset loading is read-dominated.

\paragraph{POSIX compliance gaps.}
\daxfs does not support \texttt{mknod} (device nodes, FIFOs, sockets)
or extended attributes.  File names are limited to 255 characters.
These restrictions reflect the target workloads (container rootfs,
shared caching) where these features are rarely needed.

\section{Conclusion}
\label{sec:conclusion}

CXL shared memory enables a new filesystem design point: multiple
hosts sharing a single namespace and cache through coherent load/store
operations, with no network protocol overhead.  \daxfs exploits this
by using \texttt{cmpxchg} as its sole coordination primitive,
achieving lock-free concurrent writes, cooperative caching with
decentralized MH-clock eviction, and zero-copy access in a unified
design.

We validate multi-host correctness using QEMU-emulated CXL~3.0,
confirming $>$99\% CAS accuracy under cross-host contention with no
lost updates.  On single-host DRAM-backed DAX, \daxfs exceeds tmpfs
throughput across all write workloads, achieving up to
$2.68\times$ higher random write throughput with 4 threads and
$1.18\times$ higher random read throughput at 64\,KB.  Preliminary
GPU microbenchmarks further confirm that the \texttt{cmpxchg}-based
design extends to GPU threads at PCIe~5.0 bandwidth limits.  As
CXL~3.0 hardware matures, \daxfs provides a ready filesystem layer
for disaggregated memory pools shared across hosts.

\section*{Availability}

\daxfs is available as open-source software~\cite{daxfs:repo}.  The
implementation includes the Linux kernel module, \texttt{mkdaxfs}
image creation tool, and \texttt{daxfs-inspect} debugging tool.


\end{document}